\documentclass[preprint]{acmart}
\setcopyright{none}
\renewcommand\footnotetextcopyrightpermission[1]{}
\usepackage{xspace}
\usepackage{enumitem}
\usepackage{multirow}
\usepackage{graphicx}
\usepackage{amsmath}
\usepackage{wrapfig}

\begin{document}

\title{Killing Stubborn Mutants with Symbolic Execution} 


\author{Thierry Titcheu Chekam}
\email{thierry.titcheu-chekam@uni.lu}
\orcid{0000-0002-5295-1831}
\affiliation{
  \institution{SnT, University of Luxembourg}
  \country{Luxembourg}
}

\author{Mike Papadakis}
\email{michail.papadakis@uni.lu}
\affiliation{
  \institution{SnT, University of Luxembourg}
  \country{Luxembourg}
}

\author{Maxime Cordy}
\email{maxime.cordy@uni.lu}
\affiliation{
  \institution{SnT, University of Luxembourg}
  \country{Luxembourg}
}

\author{Yves Le Traon}
\email{yves.letraon@uni.lu}
\affiliation{
  \institution{SnT, University of Luxembourg}
  \country{Luxembourg}
}

\newcommand{\tgtoolname}{{\em SEMu}\xspace}
\newcommand{\SOTAtoolname}{{\em infection-only}\xspace}
\newcommand{\changing}[1]{{\color{black} #1}} 


\begin{abstract}
We introduce \tgtoolname, a Dynamic Symbolic Execution technique that generates test inputs capable of killing stubborn mutants (killable mutants that remain undetected after a reasonable amount of testing). 
\tgtoolname aims at mutant propagation (triggering erroneous states to the program output) by incrementally searching for divergent program behaviours between the original and the mutant versions. 
We model the mutant killing problem as a symbolic execution search within a specific area in the programs' symbolic tree. In this framework, the search area is defined and controlled by parameters that allow scalable and cost-effective mutant killing.  
We integrate \tgtoolname in KLEE and experimented with Coreutils (a benchmark frequently used in symbolic execution studies). Our results show that our modelling plays an important role in mutant killing. 
Perhaps more importantly, our results also show that, within a two-hour time limit, \tgtoolname kills \changing{37}\% of the stubborn mutants, where KLEE kills none and where the mutant infection strategy (strategy suggested by previous research) kills \changing{17}\%.
\end{abstract}


\maketitle

\section{Introduction} \label{sec:intro}


Deep testing is often required in order to assess the core logic and the `critical' parts of the programs under analysis. Unfortunately, performing thorough testing is hard, tedious and time consuming. 
As a result testing the most important program parts requires substantial efforts, skills and experience. 


To deal with this issue, mutation testing aims at guiding the design of strong (likely fault revealing) test cases. The key idea of mutation 
 is to use artificially introduced defects, called mutants, to identify the weaknesses of test suites (undetected defects form test suite deficiencies) and to guide test generation (undetected defects form test objectives). Thus, testers can improve their test suites by designing test cases that take the mutation feedback into account. 

Experience with mutation testing has shown that it is relatively easy to detect a large number of mutants by simply covering them \cite{AmmannDO14, PapadakisHHJT16, PetrovicI18}. Such trivial mutants are not useful as they fail to provide any particular guidance towards test case design \cite{SchulerZ13}. However, experience has also shown that there are some few mutants that are relatively hard to detect (a.k.a. stubborn mutants \cite{YaoHJ14}) and can provide significant advantages when used as test objectives \cite{PetrovicI18, YaoHJ14}. 
 Interestingly, these mutants form special corner cases and are linked with fault revelation \cite{ChekamPTH17}. 
The importance of using the stubborn mutants as test objectives has also been underlined by several industrial studies \cite{Delgado-PerezHG18, BakerH13} including a large study with Google developers \cite{PetrovicI18}.


Stubborn mutants are hard to detect mainly due to a) the difficulty of infecting the program state (causing an erroneous program state when executing the mutation/defective point) and b) due to the masking effects that prohibit the propagation of erroneous states to the program output (aka failed error propagation \cite{AndroutsopoulosCDHH14} or coincidental correctness \cite{doi:10.1002/stvr.1696}). Either being the case, the issues linked with these mutants form corner cases which are most likely to escape testing (since stubborn mutants form small semantic deviations) \cite{PetrovicI18}. 

Killing stubborn mutants (designing test cases that reveal undetected mutants) is challenging due to the variety of the code paths, constraints and data states of the program versions (original and mutant versions) that need to be differentially analysed. The key challenge here regards the handling of the failed error propagation (masking effects), which is prevalent in stubborn mutants. Effective error propagation analysis is still an open  problem \cite{MutationSurvey, PIZZOLETO2019} as it involves state comparisons among the mutant and the original program executions that grow exponentially with the number of the involved paths (from the mutation point to the program output). 


We present \tgtoolname, an approach based on dynamic symbolic execution that generates test inputs capable of killing stubborn mutants. The particular focus of \tgtoolname is on the effective and scalable handling of mutant propagation. 
Our technique executes both the original and mutant program versions with a single symbolic execution instance, where the mutant executions are ``forked'' when reaching the mutation points. The forked execution follows the original one and compares with it. The comparisons are performed based on the involved symbolic states and related (propagation) constraints that ensure divergences. 

A key issue with both symbolic execution and mutation testing regard their scalability. To account for this problem, we develop a framework that allows defining the mutant killing problem as a search problem within a specific area around the mutation points. This area is defined by a number of parameters that control the symbolic exploration. We thus, perform a constrained symbolic exploration, starting from a pre-mutation point (a point in the symbolic tree that is before the mutation point) and ending at a post-mutation checkpoint (a point after the mutant) where we differentially compare the symbolic states of the two executions (forked and original) and generate test inputs. 

We assume the existence of program inputs that can reach the areas we are targeting. Based on these inputs, we infer preconditions (a set of consistent and simplified path conditions), which we use to constrain	 the symbolic exploration to only a subset of program paths that are relevant to the targeted mutants. To further restrict the exploration to a relevant area, we systematically analyse the symbolic tree up to a relatively small distance from the mutation point (performing a shallow propagation analysis). 

To improve the chances for propagation we also perform a deep exploration of some subtrees. Overall, by controlling the above parameters we can define strategies with trade-offs between space (cost) and deepness (effectiveness). Such strategies allow the differential exploration of promising code areas, while keeping their execution time low.

Many techniques targeting mutation-based test generation have been proposed \cite{AnandBCCCGHHMOE13, MutationSurvey}. However, most of these techniques focus on generating unit-level test suites from scratch, mainly by either covering the mutants or by causing an erroneous program state at the mutation point. However, there is no work leveraging the value of existing tests to perform deep testing by targeting stubborn mutants, which are mostly hard to propagate. Moreover, none of the available symbolic execution tools aim at generating test inputs for killing mutants.

We integrated \tgtoolname\footnote{Publicly available at https://github.com/thierry-tct/KLEE-SEMu.} into KLEE \cite{CadarDE08}. 
We evaluated \tgtoolname on \changing{47} programs from Coreutils, real-world utility programs written in C, and compare it with the mutant infection strategy, denoted as infection-only, that was proposed by previous work~\cite{pexmutator, HarmanJL11}. Our results show that \tgtoolname achieves significantly higher killing rates (approximately \changing{+37}\% and \changing{+20}\%) of stubborn mutants, for both KLEE (alone) and infection-only strategy, on the majority of the studied subjects. 

In summary, our paper makes the following contributions:




\begin{enumerate}
	\item We introduce and implement a symbolic execution technique for 
	generating tests that kill stubborn mutants. Our technique leverages existing tests in order to perform a deep and targeted test of specific code areas. 
	
	\item We model the mutant killing as a search problem within a specific area (around the mutation point). Such a modelling allows controlling the symbolic execution cost, while at the same time allows forming cost-effective heuristics. 
		
	\item We report empirical results demonstrating that \tgtoolname has a strong mutant killing ability, which is significantly superior
to KLEE and other mutation-based approaches.
\end{enumerate}

\section{Context}

Our work aims at the automatic test input generation for specific methods/components of the systems under test. Our working scenario assumes that testers have performed some basic testing and want to dig into some specific parts of the program. This is a frequent scenario used to increase confidence on the critical code parts (encode the core program logic) or on parts that testers feel uncertain. To do so, it is reasonable to use mutation testing 
by adding tests that detect the surviving mutants (mutants undetected by the existing test suite) \cite{SchulerZ13, YaoHJ14}.

We consider a mutant as detected (killed) by a test when its execution leads to different observable output from that on the original program. 
According to our scenario, the targeted mutants are those (killable) that survive a reasonable amount of testing. 
This definition depends on the amount of the performed testing; strong test suites kill more mutants than weak ones, while `adequate' test suites kill them all \cite{YaoHJ14, 0020331}. 

To adopt a baseline for basic or `reasonable amount of testing' we augment the developer test suites with KLEE. This means that the \textit{stubborn mutants are those that are killable and survive the developer and automatically generated test suites}. The surviving mutants form the objectives for our test generation technique.  
%
%
%

\subsection{Symbolic Encoding of Programs}

Independently of its language, we define a program as follows.

 \begin{definition}
  A program is a Labeled Transition System (LTS) $\mathcal{P}=(C, \mathit{c_0}, C_{out}, V, \mathit{eval_0}, T)$ where:
    \begin{itemize}
        \item $C$ is a finite set of control locations;
        \item $\mathit{c_0} \in C$ is the unique entry point (start) of the program;
        \item $C_{out} \subset C$ is the set of terminal locations of the program;
        \item $V$ is a finite set of variables;
        \item $\mathit{eval_0}$ is a predicate capturing the set of possible initial valuations of $V$;
        \item $T: C \times GC \rightarrow C$ is a deterministic transition function where each transition is labeled with a guarded command of the form $[g]f$ where $g$ is a guard condition and $f$ is a function updating valuation of variables $V$ ($GC$ is the set of labels).
    \end{itemize}
 \end{definition}
 
The LTS modelling a given program defines the set of control paths from $c_0$ to any $c_{out} \in C_{out}$. A path is a sequence of $n$ connected transitions $\pi_P = \langle (c_0, gc_0, c_1), \dots , (c_{n-1}, gc_{n-1}, c_{n = out}) \rangle$ such that $(c_i, gc_i, c_{i+1}) \in T$ for all $i$. Any well-terminating execution of the program goes through one such path. Since we consider deterministic programs, this path is unique and determined by the initial valuation (i.e. the test input) $v_0$ of the variables $V$. More precisely, each path $\pi_P$ defines a \emph{path condition} $\phi(\pi_P)$ which symbolically encodes all executions going through $\pi_P$. This path condition consists of a Boolean formula such that the test with input $v_0$ executes through $\pi_P$ iff $v_0 \models \phi(\pi_P)$. By solving $\phi(\pi_P)$ (e.g. with a constraint solver like Z3 \cite{Z3Solver}), one can obtain an initial valuation satisfying the path condition, thereby obtaining a test input that goes through the corresponding program path.

The execution of the resulting test input is a sequence of $n+1$ couples of variable valuations and locations, noted $\tau_{(P,v_0)} =  \langle (v_0, c_0)$, $\dots$, $(v_{n-1}, c_{n-1})$, $(v_{n = out}, c_{n = out})\rangle$, such that $v_0 \models eval_0$ and for all $i$, $v_i \models g_i$ and $v_{i+1} = f_i(v_i)$. While $v_{out}$ is the valuation of all variables when $\tau_{(P,v_0)}$ terminates, the observable result of $\tau_{(P,v_0)}$ (its \emph{output}), noted $Out(\tau_{(P,v_0)})$, is the subset of $v_{out}$ restricted only to all observable variables. Since a path $\pi$ encompasses a set of executions, we can also represent the set of outputs of those executions into a symbolic formula $Out(\pi)$.

\subsection{Symbolic Encoding of Mutants} \label{sec:killingamutant}

A mutation alters or deletes a statement of the original program $P$. Thus, a mutant is defined as a change in the transitions of $P$ that correspond to that statement (i.e. two transitions for branching statements; one for the others).


\begin{definition} \label{def:mutant}
	Let $\mathcal{P}=(C, \mathit{c_0}, V, \mathit{eval_0}, T)$ be an original program. A mutant of $\mathcal{P}$ is a program $\mathcal{M}=(C, \mathit{c_0}, V, \mathit{eval_0}, T')$ with $T' = (T \backslash T_m) \cup T'_m$ such that:
	\small
  \[
   \begin{cases} 
    T_m \subseteq T \land |T_m| > 0\\
    \forall (c_1, [g']f', c'_2) \in T'_m , \exists (c_1, [g]f, c_2) \in T_m : ([g']f'\neq [g]f) \lor (c'_2\neq c_2)\}
   \end{cases}
	\]
	\normalsize \
\end{definition}
  

It may happen that a program mutation leads to an \emph{equivalent} mutant (i.e. semantically equivalent to the original program), that is, for any test input $t$, $Out(\tau_{(P,v_0)}) \equiv Out(\tau_{(M,v_0)})$. All non-equivalent mutants, however, should be discriminated (i.e. \emph{killed}) by at least one test input. Thus, there must exist a test input $t$ that satisfies the following three conditions (referred to as RIP~\cite{0020331,DeMilloO91,Morell:1990:TheoryOfFaultBasedTesting}): the execution of $t$ on $M$ must (i) reach a mutated transition, (ii) infect (cause a difference in) the internal program state (i.e. change the variable valuations or the reached control locations), (iii) propagate this difference up to the program outputs. One can encode those conditions as the symbolic formula: $kill(P,M) \triangleq \exists \pi_P, \pi_M : \phi(\pi_P) \land \phi(\pi_M) \land (Out(\pi_P) \not\equiv Out(\pi_M))$. Any valuation satisfying this formula forms a test input killing $M$. For given $\pi_P$ and $\pi_M$, $kill(\pi_P,\pi_M) \triangleq \phi(\pi_P) \land \phi(\pi_M) \land (Out(\pi_P) \not\equiv Out(\pi_M))$ denotes the formula encoding the test inputs that kills $M$ and go through $\pi_P$ and $\pi_M$ in $P$ and $M$, respectively.

\begin{definition}
Let $P$ be an original program and $M_1, \dots, M_n$ be a set of mutants of $P$. Then the \textbf{mutant killing problem} is the problem of finding, for each mutant $M_i$:

\begin{enumerate}
	\item two paths $\pi_P$ and $\pi_{M_i}$ such that $kill(\pi_P, \pi_{M_i})$ is satisfiable;
	\item a test input $t$ satisfying $kill(\pi_P, \pi_{M_i})$.
\end{enumerate}
\end{definition}


\begin{figure*}[!t]
	\centering
	\vspace{-1.0em}
	\includegraphics[width=0.95\linewidth, trim={0 0.5cm 0 0.1cm}]{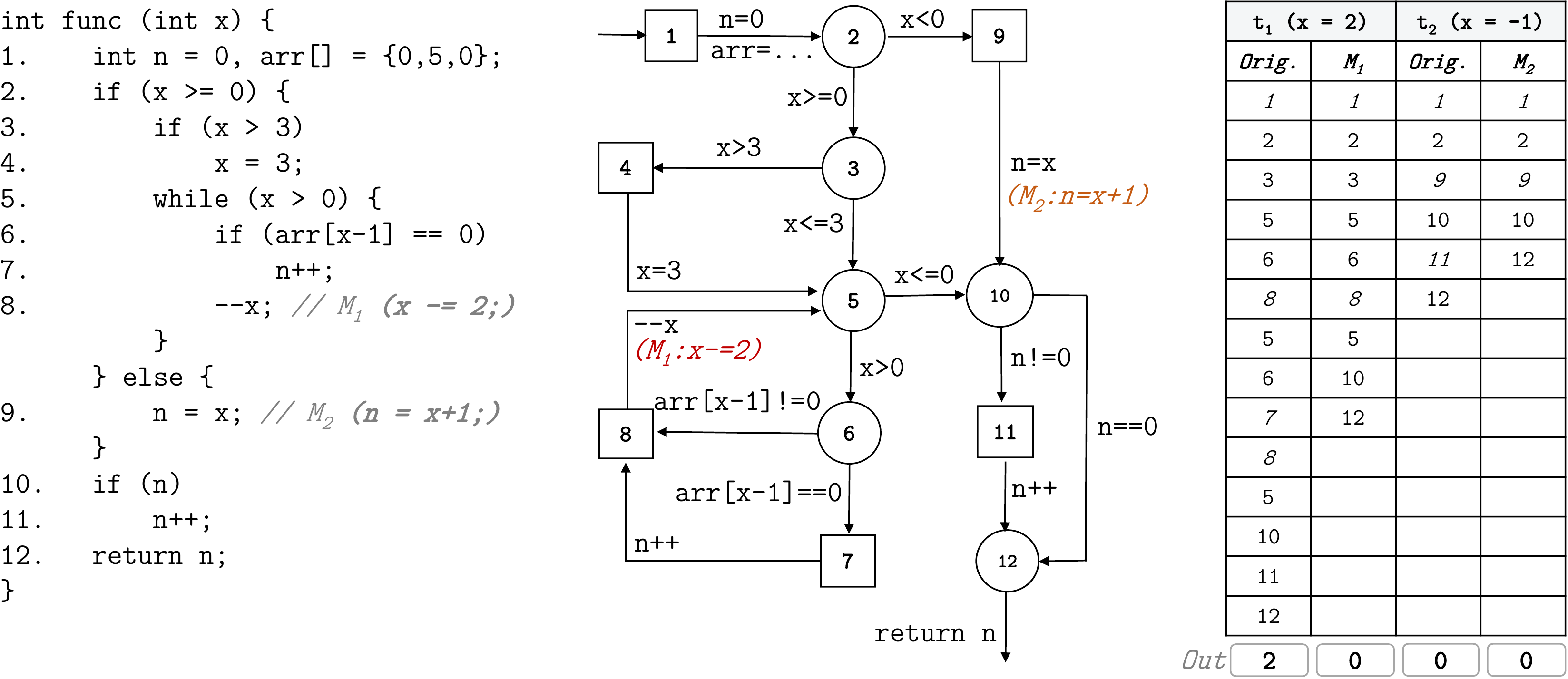}
	\caption{Example. The rounded control locations represent conditionals (at least 2 possible transition from them). }
		\label{fig:example_prog_representation}
	\vspace{-0.2em}
\end{figure*}


\subsection{Example}
Figure~\ref{fig:example_prog_representation} shows a simple C program. The corresponding C code and transition system are shown in the left and middle of Figure~\ref{fig:example_prog_representation}, respectively. The transition system does not show the guarded commands for readability. The right side of Figure~\ref{fig:example_prog_representation} shows two test inputs and their corresponding traces (as sequences of control locations of the transition system). The transition system contains $12$ control locations, corresponding to the 12 numbered lines in the program. The squared nodes of the transition system represent the non-branching control locations and the circular nodes represent the branching control location. For simplicity, we assume that each line is atomic. The initial condition $eval_0$ is $x \in Int$ where $Int$ is the set of all integers. Two mutants $M_1$ and $M_2$ are generated by mutating statements $8$ and $9$, respectively. $M_1$ results from changing the statement $``--x"$ into $``x-=2"$ and $M_2$ results from changing the statement $``n=x"$ into $``n=x+1"$. The mutants $M_1$ and $M_2$ result from the mutation of the guarded command of the transitions $8\rightarrow 5$ and $9\rightarrow 10$, respectively. 

The test execution of $t_1$ reaches $M_1$ but not $M_2$, while $t_2$ reaches $M_2$ but not $M_1$. Test $t_1$ infects $M_1$ and $t_2$ infects $M_2$. The test execution of $t_1$ on the original program and mutant $M_1$ return $2$ and $0$, respectively. The mutant $M_1$ is killed by $t_1$ because $2 \neq 0$. Similarly, the test execution of $t_2$ on the original program and mutant $M_2$ return $0$ and $0$, respectively. Test $t_2$ does not kill mutant $M_2$.

\section{Symbolic Execution} \label{sec:symbolic_execution_overview}

One can apply symbolic execution to explore the different paths, using a symbolic representation of the input domain (as opposed to concrete values) and building progressively the path conditions of the explored paths. The symbolic execution starts by setting an initial path condition to $\phi=True$. At each location, it evaluates (by calling a dedicated solver) the guarded command of any outgoing transition. If the conjunction of the guard condition and $\phi$ is satisfiable then there exists at least one concrete execution that can go through the current path and the considered transition. In this case, the symbolic execution reaches the target location and $\phi$ is updated by injecting into it the guarded command of the transition. When multiple transitions are available, the symbolic execution successively chooses one and pursues the exploration, e.g. in a breadth-first manner.


As the symbolic execution progresses, it explores additional paths. The explored paths can together be concisely represented as a tree~\cite{King:1976:Symbex} where each node is an execution state $\langle \phi, \sigma \rangle$ made of its path condition $\phi$ and symbolic program state $\sigma$ (itself constituted by the current control location -- program counter value -- and the current symbolic valuation of variables). 

Still, the tree remains too large to be explored exhaustively. Thus, one typically guides the symbolic execution to restrict the paths to explore, effectively cutting branches of the tree. 
Preconditioned symbolic execution 
attempts to reduce the path exploration space by setting the initial path condition (at the beginning of the symbolic execution) to a specific condition. This precondition restricts the symbolic execution to the subset of paths that are feasible given the precondition. The idea is to derive the preconditions from pre-existing tests (aka \emph{seeds} in the KLEE platform) that reach the particular points of interests. This allows us to provide vital guidance towards reaching the areas that should be explored symbolically, while drastically reducing the search space. In the rest of the paper, we refer to a \textit{preconditioned symbolic execution that explores the paths followed by some concrete executions as ``seeded symbolic execution''}.
 
 Overall, one can make the following steps to generate test inputs for a program $P$ via symbolic execution:
 \begin{enumerate}
 \item {\bf Precondition}: specify a logical formula over the program inputs (computed as the disjunction of the path conditions of the paths followed by the executions of the seeds) to prune out the paths that are irrelevant to the analysis.
  \item {\bf Path exploration}: explore a subset of the paths of $P$, effectively discarding infeasible paths.
  \item {\bf Test input generation}: for each feasible path $\pi_P$, solve $\phi(\pi_P)$ to generate a test input $t$ whose execution $\tau_{(P,t)}$ follows $\pi_P$.
 \end{enumerate}

\section{Killing Mutants}

\subsection{Exhaustive Exploration} \label{sec:mutant_symbex_complete}

A direct way to generate test inputs killing some given mutants (of program $P$) is to apply symbolic execution on both $P$ and the mutants, thereby obtaining their respective set of (symbolic) paths. Then, we can solve $kill(\pi_P, \pi_{M_i})$ to generate a test input that kills mutant $M_i$ and goes through $\pi_P$ in $P$ and through $\pi_{M_i}$ in $M_i$.

\begin{figure}[!t]
	\centering
	\vspace{0.5em}
	\includegraphics[width=0.65\linewidth]{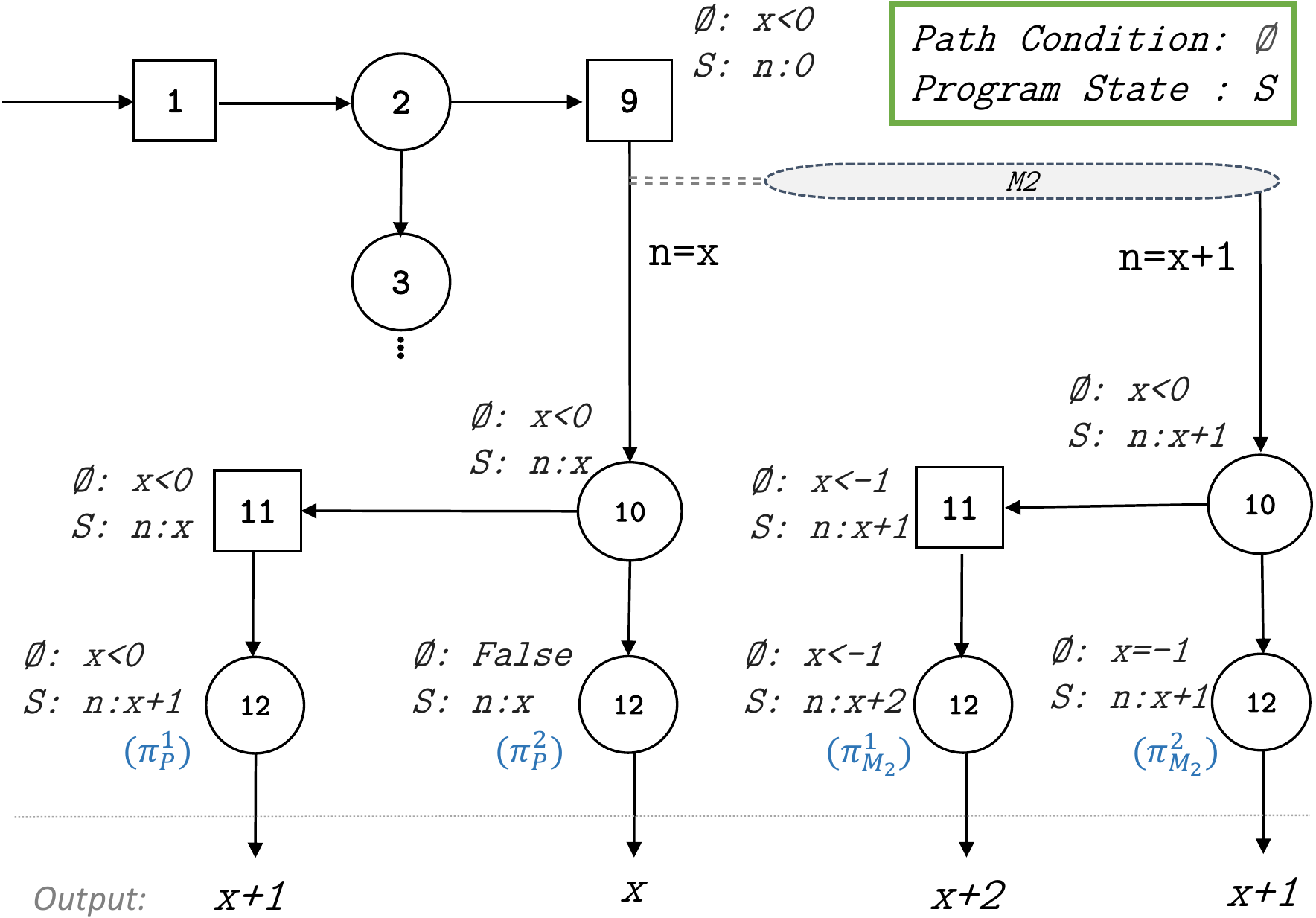}
	\caption{Example of Symbolic execution for mutant test generation. After control location $9$, the symbolic execution on the original program contains transition $9\rightarrow 10$ with $n=x$ while the symbolic execution of the mutant $M_2$ contains transition $9\rightarrow 10$ with $n=x+1$.}
		\label{fig:example_prog_mutant_symbex}
\end{figure}

Figure~\ref{fig:example_prog_mutant_symbex} illustrates the use of symbolic execution to kill mutant $M_2$ of Figure~\ref{fig:example_prog_representation}. We skip the symbolic execution subtree rooted at control location $3$ since the corresponding paths do not reach mutant $M_2$ and can easily be pruned using static analysis. Also, we do not represent the symbolic variables $arr$ and $x$, which are not updated in this example. The symbolic execution on the original program leads to the paths $\pi^1_P$ and $\pi^2_P$ such that $\phi(\pi^1_P) \equiv (x<0)$, $\phi(\pi^2_P) \equiv False$, $Out(\pi^1_P) \equiv x+1$ and $Out(\pi^2_P) \equiv x$. The symbolic execution on the mutant $M_2$ leads to the paths $\pi^1_{M_2}$ and $\pi^2_{M_2}$ such that $\phi(\pi^1_{M_2}) \equiv (x<-1)$ and $\phi(\pi^2_{M_2}) \equiv (x=-1)$, and $Out(\pi^1_{M_2}) \equiv (x+2)$ and $Out(\pi^2_{M_2}) \equiv (x+1)$.

The test generation that targets mutant $M_2$ solves the following formulae:
\begin{enumerate}
	 \item $kill(\pi^1_{P}, \pi^1_{M_2})$. Satisfiable: example solution is $x=-2$.
	 \item $kill(\pi^1_{P}, \pi^2_{M_2})$. Unsatisfiable: no possible output difference.
	 \item $kill(\pi^2_{P}, \pi^1_{M_2})$. Unsatisfiable: infeasible path ($\pi^2_{P}$).
	 \item $kill(\pi^2_{P}, \pi^2_{M_2})$. Unsatisfiable: infeasible path ($\pi^2_{P}$).
\end{enumerate}
This method effectively generates tests to kill killable mutants. However, it requires a complete symbolic execution on $P$ and on each mutant $M_i$. This implies that (i) all the path conditions and symbolic outputs have to be stored and analysed, and (ii) $kill(\pi_P, \pi_{M_i})$ has to be solved possibly for each pair of paths $(\pi_P, \pi_{M_i})$. This leads to large computational cost that makes the approach impractical. 

\subsection{Conservative Pruning of the Search Space}

To reduce the computational costs induced by the exhaustive exploration we apply two safe optimizations (preserve all opportunities to kill the mutants) that prune the space of program paths. We take advantage of the fact that mutants are simple syntactic alterations that share a large portion of their code with the original program. 

\subsubsection{Meta-mutation} \label{sec:metamutation}

Our first optimization stems from the observation that all paths and path prefixes of the original program $P$ that do not include a mutated statement (i.e. location whose outgoing transitions have changed in the mutants) also belong to the mutants. Thus, the symbolic execution of $P$ and that of the mutants may explore a significant number of identical path prefixes. As seen in Figure~\ref{fig:example_prog_mutant_symbex}, the symbolic execution is identical for the original and mutant $M_2$ up to control location $9$. Instead of making two separate symbolic executions, \tgtoolname performs a shared symbolic execution based on a meta-mutant program. A meta-mutant~\cite{UntchOH93, PapadakisM10, PapadakisM11} represents all mutants in a single code. A branching statement (named \emph{mutant choice statement}) is inserted at each mutation point and controls, based on the value of a special global variable (the mutant ID), the execution of the original and mutant programs. 

The symbolic execution on the meta-mutant program initialises the mutant ID to an unknown value and explores a path normally until it encounters a mutant choice statement. Then, the path is duplicated once for the original program and once for each mutant, with the mutant ID set to the corresponding value, and each duplicated path is further explored normally. While the effect of this optimization is limited to the prefixes common to the program and all mutants, it reduces the overall cost of exploration at insignificant computation costs and without compromising the results.



\subsubsection{Discarding non-infected mutant paths}

In practice, many execution paths reach (cover) a mutant but fail to infect the program state (introducing an erroneous program state). Extending the execution along such paths is a waste of effort as the mutant will not be killed along those paths. Thus, \tgtoolname terminates anticipatively the exploration of any path that reaches the mutant but fails to infect the program state.

\subsection{Heuristic Search}

Even with the aforementioned optimizations, the exhaustive exploration procedure remains too costly due to two factors: the size of the tree to explore and the number of couples of paths $\pi_P$ and $\pi_{M_i}$ to consider. To speed up the analysis, one can further prune the search space, at the risk of generating useless test inputs (that kill no mutant) or missing opportunities to kill mutants (by ignoring relevant paths).

A first family of heuristics reduce the number of paths to explore by selecting and prioritizing them, at the risk of discarding paths that would lead to killing mutants. A second family stop exploring a path after $k$ transitions and solve, instead of $kill(\pi_P, \pi_{M_i})$, the formula  
\begin{multline*} partialKill(\pi_P[..k], \pi_{M_i}[..k]) \triangleq  \phi(\pi_P[..k]) \land \phi(\pi_{M_i}[..k]) \land (\sigma(\pi_{P}[..k]) \neq \sigma(\pi_{M_i}[..k])) \end{multline*} 
where, for any path $\pi$, $\pi[..k]$ denotes the prefix of $\pi$ of length $k$ and where $\sigma(\pi[..k])$ is the symbolic state reached after executing $\pi[..k]$. It holds that $kill(\pi_P, \pi_{M_i})$ $\Rightarrow$ $\exists k : partialKill(\pi_P[..k]$, $\pi_{M_i}[..k])$, since a mutation cannot propagate to the output of the program if it does not infect the program in the first place. The converse does not hold, though: statements after the mutation can cancel the effects of an infection, rendering the output unchanged at the end of the execution. The problem then boils down to selecting an appropriate length $k$ where to stop the exploration, so as to maximize the chances of finding an infection that propagates up to the observable outputs.




As illustrated in Figure~\ref{fig:example_prog_mutant_symbex}, generating a test at $k=3$ (control location $10$), requires to solve the constraint $partialKill(\pi_P[..3], \pi_M[..3])$ $\equiv$ $(x<0\land x\neq x+1)$. The constraint solver may return $x=-1$ which does not propagate the infection to the output. However, generating a test at $k=5$ (control location $12$), using the original path $\pi^1_P$ and mutant path $\pi^1_{M_2}$, requires to solve the constraint $x<0 \land x<-1 \land (x+1\neq x+2)$. Any value returned by the constraint solver kills the mutant.

An ideal method to kill a mutant $M$ would explore only one path $\pi_P$ and one path $\pi_M$, and up to the smallest prefix length $k$ where the constraint solver can generate a test that kills $M$. 
However, identifying the right $\pi_M$ and the optimal $k$ is hard, as it requires precisely capturing the program semantics. To overcome this difficulty, \tgtoolname defines heuristics to prune non-promising paths on the fly and to control at what point (what prefix length $k$) to call the constraint solver. Once candidate path prefixes are identified, \tgtoolname invokes the solver to solve $partialKill(\pi_P[..k], \pi_M[..k])$.





\section{\tgtoolname Cost-Control Heuristics}
\label{sec:parameters_definition}

\tgtoolname consists of parametric heuristics to control the symbolic exploration of promising code regions. Any configuration of \tgtoolname sets the parameters of the heuristics, which together define which paths to explore and the test generation process. \tgtoolname also takes as inputs the original program, the mutants to kill and a set of pre-existing test inputs to drive the seeded symbolic execution. During the symbolic exploration, \tgtoolname selects which paths to explore and when to stop the exploration to generate test inputs based on the obtained path prefix.


\begin{figure}[!t]
	\centering
	\vspace{1.5em}
	\includegraphics[width=0.99\linewidth]{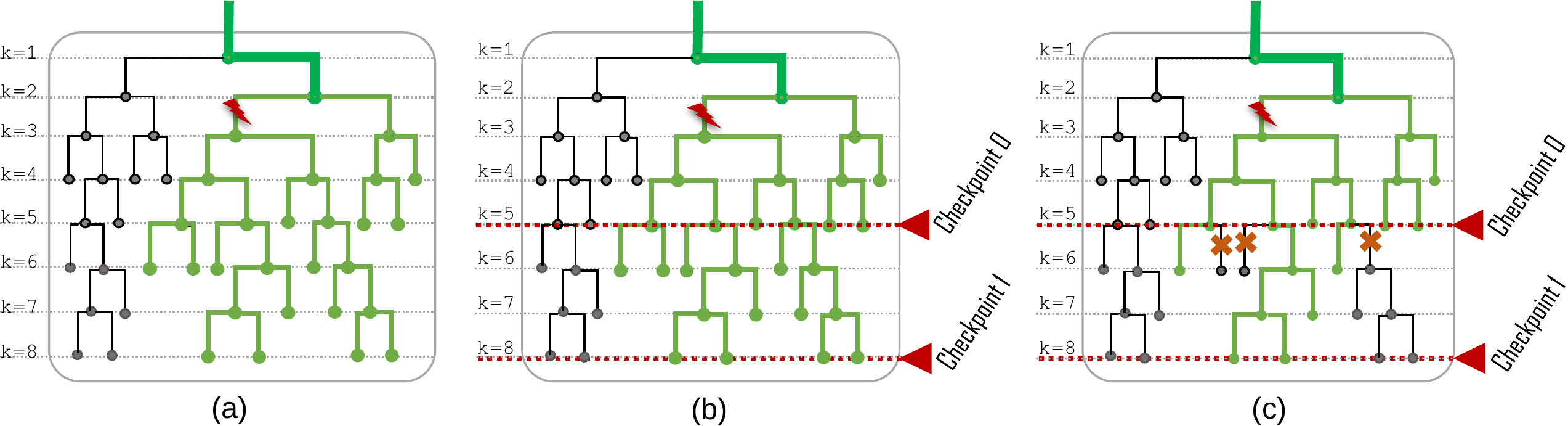}
	\caption{Illustration of \tgtoolname cost-control parameters. Subfigure (a) illustrates the Precondition Length where the green subtree represents the candidate paths constrained by the precondition (the thick green path prefix is explored using seeded symbolic execution). Subfigure (b) illustrates the Checkpoint Window (here CW is 2). Subfigure (c) illustrates the Propagation Proportion (here PP is 0.5) and the Minimum Propagation Depth (here if MPD is 1 the first test is generated, for unterminated paths, from \emph{Checkpoint 1}).}

		\label{fig:example_prog_params}
\end{figure}



\subsection{Pre Mutation Point: Controlling for Reachability} 

To improve the efficiency of the path exploration, it is important to quickly prune paths that are infeasible (cannot be executed) or irrelevant (cannot reach the mutants). To achieve this, we leverage seeded symbolic execution (as implemented in KLEE) where the seeds are pre-existing tests. 
We proceed in two steps. First, we explore the paths in seeded mode up to a given length (precondition length). Then, we stop following the seeds' executions and proceed with a non-seeded symbolic execution. The location of the switching point thus determines where the exploration stops using the precondition. In particular, if it is set to the entry point of the program then the execution is equivalent to a full non-seeded symbolic execution. If it is set beyond the output then it is equivalent to a fully seeded symbolic execution. Formally, let $\Pi$ denote the complete set of paths of a program $P$, $\{t_1, \dots, t_n\}$ be the set of seeds, and $l$ be the chosen precondition length. Then the sets of explored paths resulting from the seeded symbolic execution of length $l$ and with seeds $\{t_1, \dots, t_n\}$ is the largest set $\Pi' \subseteq \Pi$ satisfying
$\pi \in \Pi' \Rightarrow \exists t_i : t_i \models \phi(\pi[..l]).$

This heuristics is illustrated in Figure~\ref{fig:example_prog_params}a where the thick (green) segments represent the portion of the tree explored by seeded symbolic execution and the subtree below (light green) represents the portion explored by non-seeded symbolic execution. The precondition leads to pruning the leftmost subtree.

Accordingly, the first parameter of \tgtoolname controls the \emph{precondition length (PL)} at which to stop the seeded symbolic execution. Instead of demanding a specific length $l$, the parameter can take two values reflecting two strategies to define $l$ dynamically: \emph{GMD2MS} (Global Minimum Distance to Mutant Statement) and \emph{SMD2MS} (Specific Minimum Distance to Mutant Statement). When set to GMD2MS, the precondition length is defined, \emph{for all explored paths}, as the length of the smallest path prefix that reaches a mutated statement. When set to SMD2MS, the precondition length is defined, \emph{individually for each path $\pi$}, as the length $l$ of the smallest prefix $\pi[..l]$ of this path that reaches a mutated statement.

\subsection{Post Mutation Point: Controlling for Propagation}

From the mutation point, all paths of the original program are explored. When it comes to a mutant, however, it happens that path prefixes that cover and infect the program state fail to propagate the infection to the outputs. These prefixes should be discarded to reduce the search space. Accordingly, our next set of parameters controls where to check that the propagation goes on, the number of paths to continue exploring from those checkpoints, and when to stop the exploration and generate test inputs. Overall, those parameters contribute to reducing the number of paths explored by the symbolic execution as well as the length $k$ of the path prefixes from which tests are generated.

\subsubsection{Checkpoint Location}

The first parameter is an integer named the \emph{Checkpoint Window} (CW) which determines the location of the checkpoints. Any checkpoint is a program location with branching statements (i.e. transitions with guarded command $[g]f$ such that $g \neq True$) that is found after the mutation point. Then, the checkpoint window defines the number of branching statements (that are not checkpoints) between the mutation point and the first checkpoint, and between any two consecutive checkpoints. The effect of this parameter is illustrated in Figure~\ref{fig:example_prog_params}b. The marked horizontal lines represent the checkpoints. In this case, the checkpoint window is set to 2, meaning that there are two branching statements between two checkpoints. At each checkpoint, \tgtoolname can perform two actions: (1) discard some branches (path suffixes) of the current path prefix (by ignoring some of the branches) and (2) generate tests based on the current prefix. Whether and how those two actions are performed is determined according to the following parameters.


\subsubsection{Path Selection}


The parameter \emph{Propagating Proportion} (PP) specifies the percentage of the branches that are kept to pursue the exploration, whereas the parameter \emph{Propagation Selection Strategy} (PSS) determines the strategy used to select these branches. We implemented two strategies: random (RND) and Minimum Distance to Output (MDO). The first one simply selects the branches randomly with a uniform probability. The second one assigns a higher priority to the branches that can lead to the program output more rapidly (i.e. by executing fewer statements). This distance is estimated statically based on the control flow and call graphs of the program. The two parameters are illustrated in Figure~\ref{fig:example_prog_params}c, where the crossed subtrees represent branches pruned at Checkpoint 0.




\subsubsection{Early Test Generation}

Generating test inputs before the end of the symbolic execution (on the path prefixes) allows us to reduce its computation cost. Being placed after the mutation point, all checkpoints are potential places where to trigger the test generation. However, generating sooner reduces the chances of seeing the infection propagate to the program output. To alleviate this risk, we introduce the parameter \emph{Minimum Propagation Depth} (MDP), which specifies the number of checkpoints that the execution must pass through before starting to generate tests. In Figure \ref{fig:example_prog_params}c, if MDP is set to 1 then tests are generated from Checkpoint 1 (for the two remaining paths prefixes). Note that in case MDP is set to 0, tests are generated for the crossed (pruned) path prefixes at Checkpoint 0.



\subsection{Controlling the Cost of Constraint Solving}

Remember that $partialKill$ requires the state of the original program and the mutant to be different. The subformulae representing the symbolic program states can be large and/or complex, which may hinder the performance of the invoked constraint solver. To reduce this cost, we devise a parameter \emph{No State Difference} (NSD) that determines whether to consider the program state differences when generating tests. When set to $True$, $partialKill(\pi_P[..k], \pi_M[..k])$ is reduced to $\phi(\pi_P[..k]) \land \phi(\pi_M[..k])$; however, its solution has lower chances of killing mutant $M$.


\subsection{Controlling the Number of Attempts}

It is usually sufficient to generate a single test that covers the mutant to kill it. However, the stubborn mutants that we target may not be killed by the early attempts (applied closer to the mutation point) and require deeper analysis. Furthermore, a test generated to kill a mutant may collaterally kill another mutant. For those reasons, generating more than one test for a given mutant can be beneficial. Doing this, however, comes at higher test generation and test execution costs. To control this, we devise a parameter \emph{Number of Tests Per Mutant} (NTPM) that specifies the number of tests generated for each mutant (i.e. the number of $partialKill$ formulas solved for each mutant).


\section{Empirical Evaluation} \label{sec:evaluation}
\subsection{Research Questions}

We first empirically evaluate the ability of \tgtoolname to kill stubborn mutants. This is an essential question, since there is no point in evaluating \tgtoolname if it cannot kill some of the targeted mutants. 

\begin{description}
    \item[RQ1] What is the ability of \tgtoolname to kill stubborn mutants?
\end{description}



Since the results of RQ1 indicate a strong killing ability of \tgtoolname, we turn our attention to the question of whether the killing ability is due to the extended symbolic exploration that is anyway performed by KLEE. We thus, compare \tgtoolname with KLEE by running KLEE in the seed mode (using the initial test suite as a seed for KLEE test generation) to generate additional tests. Such a comparison is also a first natural baseline to compare with. These motivate RQ2:

\begin{description}
    \item[RQ2]  How does \tgtoolname compare with KLEE in terms of killed stubborn mutants?
\end{description}

Perhaps not surprisingly, we found that \tgtoolname outperforms KLEE. This provides evidence that our dedicated mutation-based approach is indeed suitable for mutation-based test generation. At the same time though, our results raises further questions on whether the superior killing ability of \tgtoolname is due to mutant infection (suggested by previous research) or due to mutant propagation (specific target of \tgtoolname). In case we find that mutant infection is sufficient for killing stubborn mutants then mutant propagation should be skipped in order to save effort and resources. To investigate this, we ask:

\begin{description}
    \item[RQ3]  How does \tgtoolname compare with the infection-only strategy in terms of killed stubborn mutants?
\end{description}

\subsection{Test Subjects}
To answer our research questions, we experimented with the C programs of GNU Coreutils\footnote{https://www.gnu.org/software/coreutils/} (version 8.22). GNU Coreutils is a collection of text, file, and shell utility programs widely used in unix systems. The whole codebase of Coreutils is made of more than 60,000 lines of C code\footnote{Measured with cloc (http://cloc.sourceforge.net/)\label{foot:cloc}}.

The repository of Coreutils contains developer tests for the utilities programs which are system tests written in shell or perl scripts that involve more than 20,000 lines of code\textsuperscript{\ref{foot:cloc}}.

Applying mutation analysis on all Coreutils programs requires excessive amount of effort. Therefore, we randomly sampled \changing{60} programs, based on which we performed our analysis. Unfortunately, in \changing{13} of them mutation analysis took excessive computational time (due to costly test execution), for which we terminated the analysis. Therefore, we analysed \changing{47} programs. These are: \texttt{\changing{\small{base64, basename, chcon, chgrp, chmod, chown, chroot, cksum, date, df, dirname, echo, expr, factor, false, groups, join, link, logname, ls, md5sum, mkdir, mkfifo, mknod, mktemp, nproc, numfmt, pathchk, printf, pwd, realpath, rmdir, sha256sum, sha512sum, sleep, stdbuf, sum, sync, tee, touch, truncate, tty, uname, uptime, users, wc, whoami}}}. The following Figure 
presents the size of these subjects. 
\\
\begin{wrapfigure}{r}{0.9\textwidth}
\vspace{-2.5em}
  \begin{center}
    \includegraphics[width=0.88\linewidth]{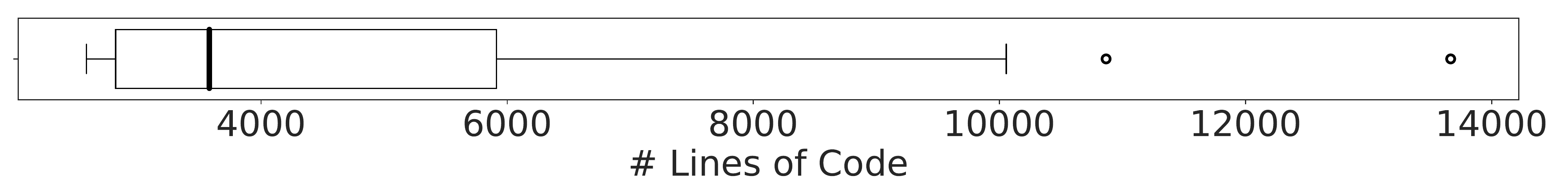}
  \end{center}\vspace{-1.5em}
\end{wrapfigure}
\\
\\
\\
For each subject we selected the 3 functions that were covered by the largest number of developer tests (from the initial test suite).

\subsection{Employed Tools} \label{sec:tool_and_configs}

We implemented our approach on top of LLVM\footnote{https://llvm.org/} using the symbolic virtual machine KLEE \cite{CadarDE08}. The version of our tool is based on the KLEE revision \texttt{74c6155}, LLVM 3.4.2. Our implementation modified (or added) more than 8,000 lines of code on KLEE, and is publicly available\footnote{https://github.com/thierry-tct/KLEE-SEMu}. We are planning to add support for newer versions of LLVM.
To convert system tests into the format of seeds required by KLEE for the seeded symbolic execution, we use \textsc{Shadow}~\cite{PalikarevaKC16}. 

Our tool requires the targeted mutants to be represented in a meta-mutant program (presented in Section~\ref{sec:metamutation}), which were produced using the \emph{Mart}~\cite{ChekamPT19} 
mutant generation tool. \emph{Mart} mutates a program by applying a set of mutation operators (code transformations) to the original LLVM bitcode program. 




\subsection{Experimental Setup}

\subsubsection{Selected Mutants}

To perform our experiment we need to form our target mutant set. To do so, we employed \emph{Mart} by using its default configuration and generated \changing{172,919} mutants. This configuration generates a comprehensive set of mutants based on a large set of mutation operators, consisting of 816 code transformations. It is noted that the operator set includes the classical 5 operators \cite{OffuttLRUZ96} that are used by most of the todays' studies and mutation testing tools. Unfortunately, space constraints prohibit us from detailing the operator set. The interested reader is refered to \emph{Mart}'s paper for further details \cite{ChekamPT19}.

To identify the stubborn mutant set we started by eliminating trivial equivalent and duplicated mutants, and form our initial mutant set $M_1$. To do so, we applied Trivial Compiler Equivalence (TCE) \cite{PapadakisJHT15}, a technique that statically removes a large number of mutant equivalences. In our experiment, TCE removed a total number of \changing{102,612} mutants as being equivalent or duplicated. This gave us \changing{70,307} mutants to be used for our initial mutant set, i.e., $M_1$=\changing{70,307}. 

Then, we constructed our initial test suites $TS$ (composed of the developer test suite augmented with a simple test generation run of KLEE). To generate these tests with KLEE, we set a test generation timeout of 24 hours, while using the same configurations presented by the authors of KLEE~\cite{CadarDE08} (except for larger memory limit and \texttt{max-instruction-time}, set to 9GB and 30s respectively). This run resulted in \changing{5,161} tests (\changing{2,693} developer tests and \changing{2,468} tests generated by the initial run of KLEE). 

We then executed the initial test suites ($TS$) with the initial mutant set ($M_1$) and identified the live and killed mutants. The killed mutants were discarded, while the live ones formed our target mutant set (denoted it as $M_2$), i.e., $M_2$ is the target of \tgtoolname. In our experiment we found that $M_2$ included \changing{26,278} mutants, which is approximately \changing{37}\% of $M_1$. 
It is noted that $M_2$ is a superset of the stubborn mutants as it includes both stubborn and equivalent mutants. Unfortunately, judging mutant equivalence is undecidable and thus, we cannot remove such mutants before test generation. Therefore, to preserve realistic settings we are forced to run \tgtoolname on all $M_2$ mutants. 

To evaluate \tgtoolname effectiveness we need to measure the extent to which it can kill stubborn mutants. Unfortunately, $M_2$ contains a large proportion of equivalent mutants \cite{SchulerZ13}, which may result in significant underestimations of test effectiveness \cite{KurtzAOK16}. Additionally, $M_2$  may contain a large portion of subsumed mutants (mutants killed collaterally by tests designed to kill other mutants), which may inflate (overestimate) test effectiveness \cite{PapadakisHHJT16}. Although we discarded easy-to-kill mutants, it is still likely that a significant amount of `noise' still remains. 

To reduce such biases (both under and over estimations) \cite{PapadakisHHJT16, KurtzAOK16}, there is a need to filter out the subsumed mutants by forming the subsuming mutant set \cite{MutationSurvey, KintisPM10}. The subsuming mutants are mainly distinct (in the sense that killing one of them does not alter, increase or decrease, the chances of killing the others) providing objective estimations of test effectiveness. Unfortunately, identifying subsuming mutants is undecidable and thus, several mutation testers, e.g., Ammann et al. \cite{AmmannDO14}, Papadakis et al. \cite{PapadakisHHJT16}, Kurtz et al. \cite{KurtzAOK16} suggested approximating them through strong test suites. Therefore, to approximate them, we used the combined test suite that merges all tests generated by KLEE and \tgtoolname across the execution of its 128 different configurations, $\bigcup\limits_{\forall i}^{} TS_{x_i}$, where $x_0$ is KLEE and $x_i\ (i>0)$ are the \tgtoolname configurations (refer to Section \ref{configuratios} for details). This process was applied on $M_2$ and resulted in a set of \changing{529} mutants. In the rest of the paper we call the mutants belonging to $M_3$ as reference mutants. We use $M_3$ for our effectiveness evaluation. 

Overall, through our experiments we used two distinct mutant sets,  $M_2$ and $M_3$. To preserve realistic settings, the former is used for test generation, while the later is used for test evaluation (to reduce bias).

\subsubsection{\tgtoolname Configuration}
\label{configuratios}
To specify relevant values for our modelling parameters we performed ad-hoc exploratory analysis on some small program functions. Based on this analysis we specify 2 relevant values for each of the 7 parameters (defined in Section~\ref{sec:parameters_definition}). These values provided us the basis for constructing a set of configurations (parameter combinations) to experiment with. In particular the values we used are the following:
Precondition Length: \textit{GMD2MS and SMD2MS}, Checkpoint Window: \textit{0 and 3}, Propagating Proportion: \textit{0 and 0.25}, Propagating Selection Strategy: \textit{RND and MDO}, Minimum Propagation Depth: \textit{0 and 2},  
No State Difference: \textit{True and False}, Number of Tests Per Mutant: \textit{1 and 5}.


We then experiment with them in order to select the \tgtoolname configuration and form our approach. It is noted that different values and combinations form different strategies. Examining them is a non-trivial task since the number of configurations is exponentially increased, i.e., $2^7=128$ and mutant execution takes considerable amount of time. 
In our study, the total test generation of the various configurations and KLEE took roughly \changing{276} CPU days, while the execution of the mutants took approximately \changing{1,400} CPU days.  

To identify and select the most prominent configuration, we executed our framework on all test subjects under all configurations $x_i$ where $i\in[1,128]$. We restrict the symbolic execution time to 2 hours. We then randomly split the set of test subjects into 5 buckets of equal size (each one containing 20\% of the test subjects). Then, we pick 4 buckets (80\% of the test subjects) and select the best configuration by computing the ratio of killed reference mutants. We assess the generalization of this configuration on the left out bucket (5th bucket that includes 20\% of the test subjects). To reduce the influence of random effects, we repeated this process 5 times by leaving every bucket out for evaluation. At the end we selected the median performing configuration (performance on the bucket that had been left out). It is noted that such a cross validation process is commonly used in order to select stable and potentially generalizable configurations. 

Based on the above procedure we selected the \tgtoolname configuration: \changing{$PL=\mbox{GMD2MS},\ CW=0,\ PP=0.25,\ PSS=\mbox{RND},\ MPD=2,\ NSD=False,\ NTPM=5$}.

%


\subsection{Experimental Settings and Procedure} 

To perform our experiment we set, on KLEE, the following (main) settings (which are similar to the default parameters of KLEE): a) we set a memory usage threshold of 8 GB, (a threshold never reached by any of the studied methods), b) we set the search strategy on Breadth-First Search (BFS), which is commonly used in patch testing studies \cite{PalikarevaKC16} and c) we set a 2 hours time limit for each subject. 

It is noted that our current implementation supports only BFS. We believe that such a strategy fits well with our purpose as it is important that the mutants and original program paths are explored in a lock step in order to enable state comparison at the same depth.  
%
%
The time budget of 2 hours was adopted because it is frequently used in test generation studies, e.g., \cite{PalikarevaKC16}, and forms a time budget that is neither too big nor too small. It is noted that since \tgtoolname performs a deeper analysis than the other methods, adopting a higher time limit would probably lead to an improved performance, compared to the other methods. Of course reducing this limit could lead to reduced performance. 

We then evaluated the generated test suites by computing the ratio of reference mutants that they kill. Unfortunately, in \changing{11} among the \changing{47} test subjects we considered, none of the evaluated techniques managed to kill any mutant. This means that for these \changing{11}  subjects we approximate having 0 stubborn mutants and thus, we discarded those programs. Therefore, the following results regard the \changing{36} programs for which we could kill at least one stubborn mutant.

To answer RQ1 we compute and report the ratio of the reference mutants killed, i.e., $M_3$ set, by \tgtoolname when it targets the \changing{26,278} surviving mutants, i.e., $M_2$ set. 

To answer RQs 2 and 3 we compute and contrast the ratio of the reference mutants killed by KLEE (executed in "seeding" mode), the infection-only strategy (a strategy suggested by previous research~\cite{pexmutator, HarmanJL11}) and \tgtoolname (for fair comparison, we used the initial test suite as seeds for the three approaches). We also report and contrast the number of mutant-killing tests that were generated. Since the generated tests may include large numbers of redundant tests, i.e., a test is redundant with respect to a set of tests when it does not kill any  unique mutant compared to the mutants killed by the other tests in the set \cite{MutationSurvey}, we compare the sizes of non-redundant test sets, which we call mutant-killing test sets.  The size of these sets represents the raw number of end objectives that were successfully met by the techniques \cite{0020331, MutationSurvey}.

To compute the mutant-killing test sets we used a greedy heuristic. 
This heuristic incrementally selects the tests that kill the maximum number of mutants that were not killed by the previously selected tests. 

\subsection{Threats to Validity}

   All in all we targeted \changing{133} functions from \changing{47} programs from Coreutils. 
        This level of evidence sufficiently demonstrates the potential of our approach, but should not be considered as a general assertion of its test effectiveness. 
        
        We generated tests at the system level, relying on the developers' tests suites. We believe that this is the major advantage of our approach
because this way we focus on stubborn mutants that encode system level corner cases that are hard to reveal. Another benefit of doing so is that at this level we can reduce false alarms, experienced at unit level (feasible behaviors at unit but infeasible at system level), \cite{GrossFZ12}. Unfortunately though, this could mean that our results do not necessarily extend to unit level.        
        
        Another issue may be due to the tools and frameworks we used. Potential defects and limitations of these tools could influence our observations. To reduce this threat we used established tools, i.e., KLEE and Mart, that have been used by many empirical studies. To reduce this threat further we also performed manual checks and made our tool publicly available. 
    
In our evaluation we used the subsuming stubborn mutants in order to cater for any bias caused by trivial mutants \cite{PapadakisHHJT16}. While this practice follows the recommendations made by the mutation testing literature \cite{MutationSurvey}, the subsuming set of mutants is a subject to the combined reference test suite, which might not be representative to the input domain. Nevertheless, any issue caused by the above approximations could only reduce the mutant killed ratios and not the superiority of our method. Additional (future) experimentations will increase the generalizability of our conclusions. 

The comparison between the studied methods (infection-only) was based on a time limit that did not include any actual mutant test execution time. This means that when reaching the time limit, we cannot know how successful (at mutant killing) the generated tests were. Additionally, we cannot perform test selection (eliminate ineffective tests) as this would require expensive mutant executions. While, it is likely that a tester would like to execute the mutants in order to perform test selection, leaving mutant execution out allows a fair comparison basis between the studied methods since mutant execution varies between the methods and heavily depends on test execution optimizations used \cite{MutationSurvey}. Nevertheless, it is unlikely that including the mutant execution would change our results since \tgtoolname generates less tests than the baselines (because it makes a deeper analysis than the baselines).

\section{Empirical Results}

\subsection{Killing ability of \tgtoolname}

To evaluate the effectiveness of \tgtoolname we run it for 2 hours per subject program and collect the generated test inputs. We then execute these inputs with the reference mutants and determine the killed ones. Interestingly \tgtoolname kills a large portion of the reference mutants. The median percentage of killed mutants is \changing{37.3}\%, indicating a strong killing ability. To kill these mutants  \tgtoolname generated  \changing{153} mutant-killing test inputs (each test kills at least one mutant that is not killed by any other test). 


\subsection{Comparing \tgtoolname with KLEE}

\begin{figure}[!t]
	\centering
	\vspace{-0.5em}
	\includegraphics[width=0.85\linewidth]{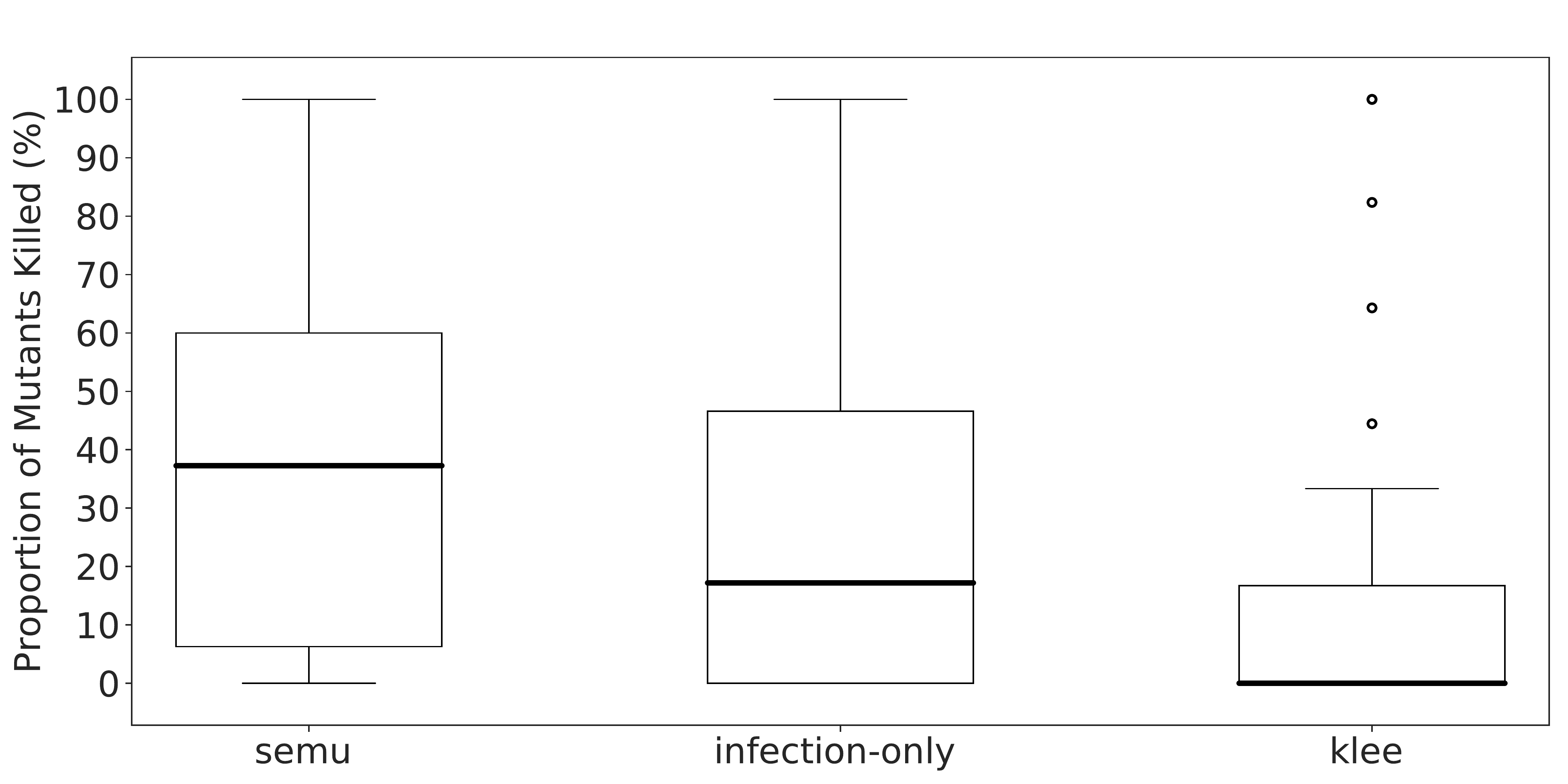}
	\caption{Comparing the stubborn mutant killing ability of \tgtoolname, KLEE and the \SOTAtoolname.}
	\vspace{-0.5em}
		\label{fig:stubborn_ms_start_comparison}
\end{figure}

Figure~\ref{fig:stubborn_ms_start_comparison} records the proportion of the killed reference mutants by \tgtoolname, seeded mode of KLEE and infection-only (investigated in RQ3). It is noted that the boxes include the proportions of killed mutants among the different test subjects we use. From these results we can observe that \tgtoolname has a median value of \changing{37.3}\% while KLEE has a median of \changing{0.0}\%. 

To further validate the difference we use the Wilcoxon statistical test (paired version) to check whether the differences are significant. The statistical test gives a p-value of \changing{0.006} suggesting that the two samples' values are indeed significantly different. As statistical significance does not provide any information related to the volume of the difference, we also compute the Vargha Delaney effect size ($\hat{A_{12}}$ value) that quantifies the frequency the observed difference. The results give a $\hat{A_{12}}$ of \changing{0.736}, which indicates that \tgtoolname is superior to KLEE in \changing{73.6}\% of the cases.

\begin{figure}[!t]
	\centering
	\includegraphics[width=0.85\linewidth, trim={0 0.52cm 0 0.15cm},clip]{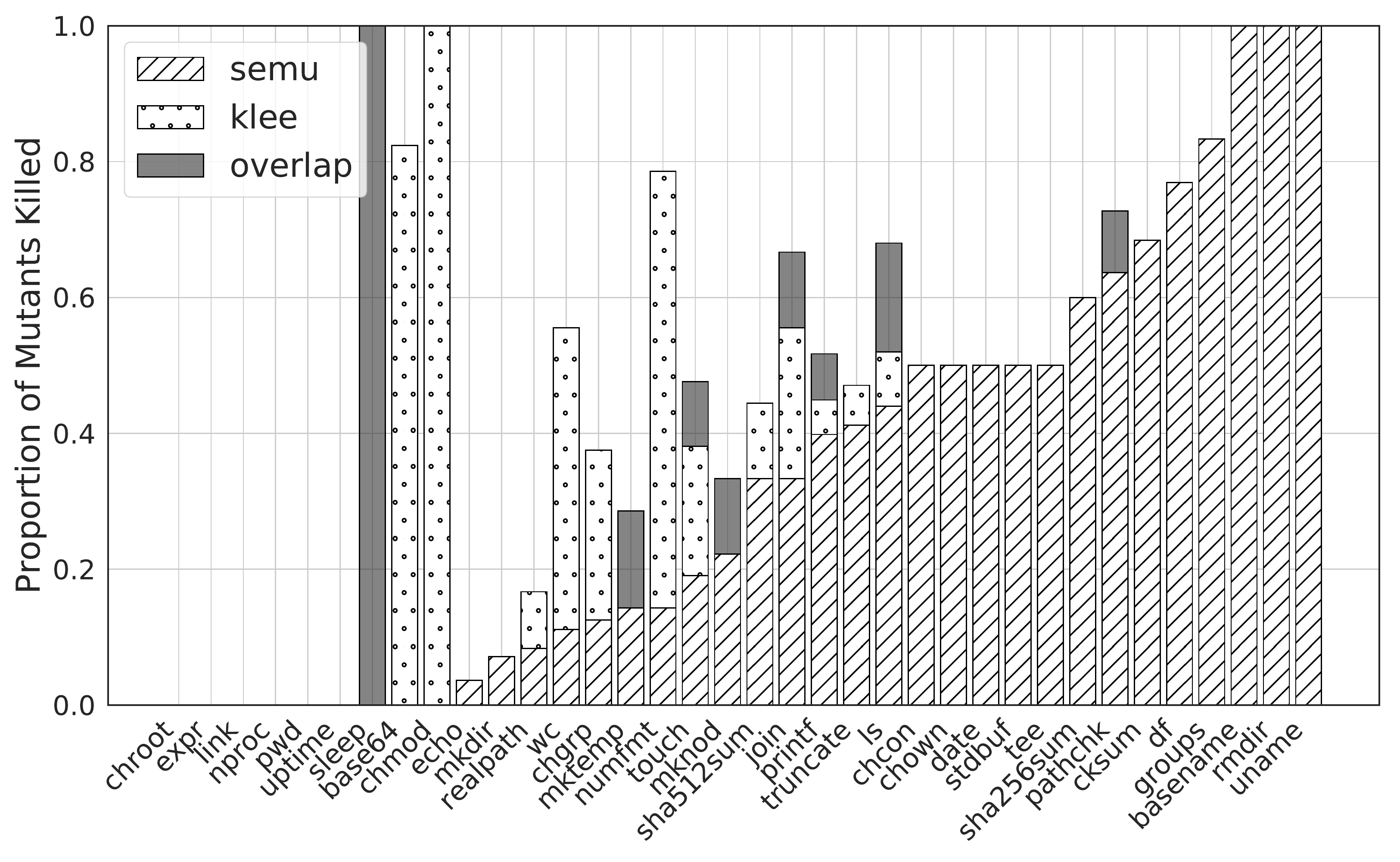}
	\caption{Comparing the mutant killing ability of \tgtoolname and KLEE in per program basis.}
		\label{fig:stubborn_overlap_semu_klee}
\end{figure}

Figure~\ref{fig:stubborn_overlap_semu_klee} depicts the differences and overlap between the reference mutants killed by \tgtoolname and KLEE, per studied subject. From this figure, we can observe that the number of programs with overlapping killed mutant is very small indicating that the two methods differ significantly. We also observe that \tgtoolname performs best in the majority of the cases. Interestingly, a non negligible number of mutants are killed by KLEE only. These cases fall within a small number of test subjects. We investigated these cases and found that the differences were big either because there was only one reference mutant, which was killed by KLEE alone, or because of the large number of surviving mutants that force \tgtoolname perform a shallow search. Unfortunately, \tgtoolname spends much time trying to kill every targeted mutant and thus, when a large number of them is involved, the 2 hours time limit we set is not sufficient to effectively kill them.

To better demonstrate the effectiveness differences of the methods we also record the number of the mutant killing test inputs (each test kills at least one mutant that is not killed by any other test). We found that \tgtoolname generated \changing{153} mutant-killing test inputs, while KLEE generated only \changing{62}.

\subsection{Comparing \tgtoolname with infection-only}

A first comparison between \tgtoolname and \SOTAtoolname can be made based on the data from Figure~\ref{fig:stubborn_ms_start_comparison}. According to these data \tgtoolname has a median value of \changing{37.3}\% while \SOTAtoolname has a median of \changing{17.2}\%. Interestingly, this shows a big difference in favour of our approach. To further validate this finding, we performed a Wilcoxon statistical test and got a p-value of \changing{0.04} suggesting that the two samples' values are statistically significant (at the commonly adopted 5\% confidence level). Like in RQ2 we also computed the Vargha Delaney effect size $\hat{A_{12}}$ and found that \tgtoolname yields higher killing rates than \SOTAtoolname in \changing{61}\% of the cases. 

\begin{figure}[!t]
	\centering
	\includegraphics[width=0.85\linewidth, trim={0 0.01cm 0 0.01cm},clip]{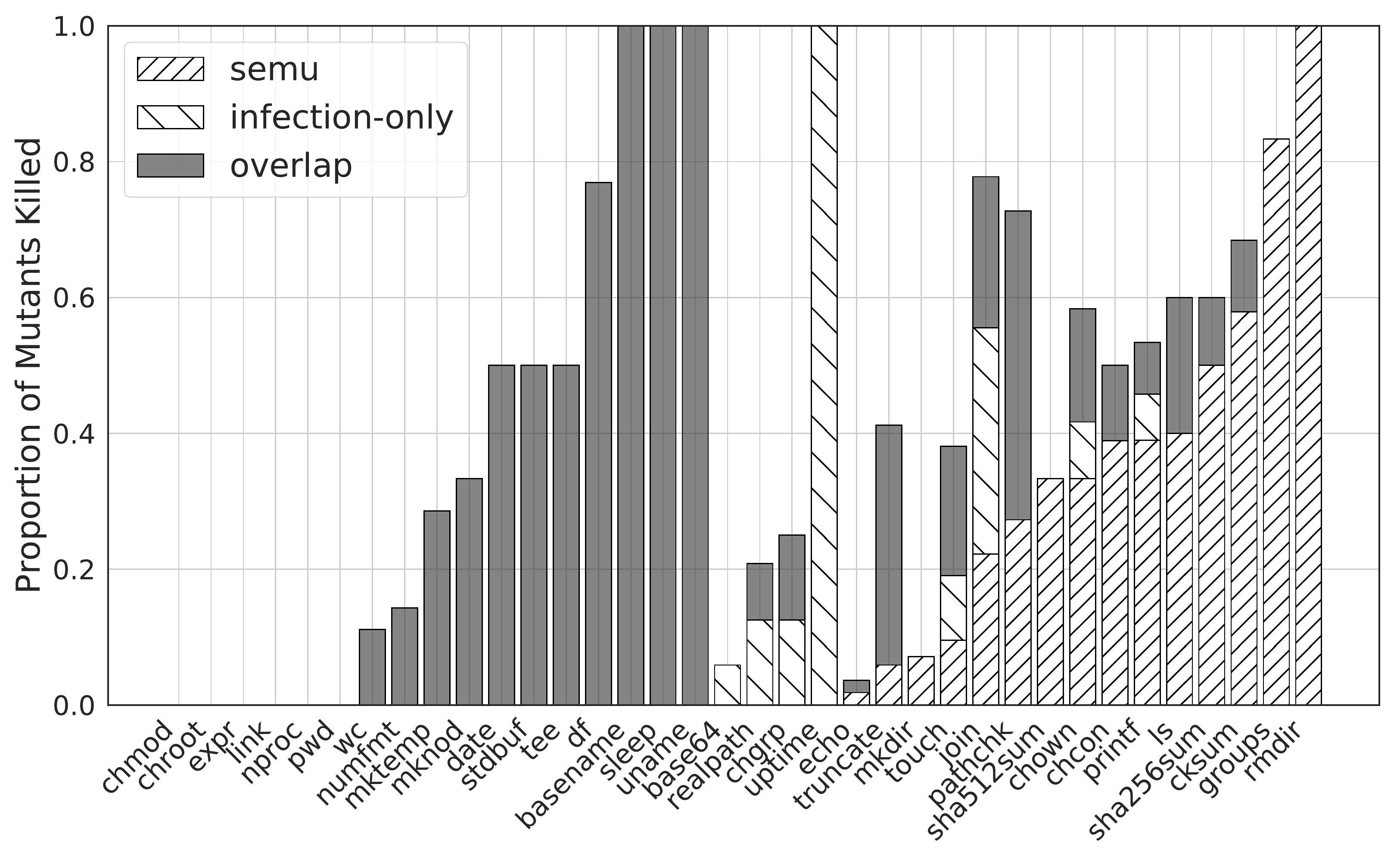}
	\caption{Comparing the mutant killing ability of \tgtoolname and \SOTAtoolname in per program basis.}
		\label{fig:stubborn_overlap_semu_sota}
\end{figure}

To demonstrate the differences we also present our results in a per test subject basis. Figure~\ref{fig:stubborn_overlap_semu_sota} shows the differences and overlap between the killed reference mutants. From these results we observe a large overlap between the mutants killed by both approaches, with \tgtoolname being able to kill more mutants for most of the cases. We also observe that in \changing{5} of the cases \SOTAtoolname performed better than \tgtoolname, while \tgtoolname performed better in \changing{13}. 

Similarly, to the previous RQs we compare the strategies by counting the number of the mutant killing test inputs that were generated by the strategies.  Interestingly, we found that \tgtoolname generated \changing{87}\% more mutant killing test inputs than the  "infection-only" one (\changing{153} vs. \changing{82} inputs) , indicating the usefulness of our framework.

\section{Related Work}  \label{sec:related_works}

Many techniques targeting mutation-based test generation have been proposed \cite{AnandBCCCGHHMOE13, MutationSurvey}.  However, most of these  focus on generating test suites from scratch, by maximizing the number of mutants killed, mainly by either covering the mutants or by targeting mutant infection. In contrast we aim at deep testing of specific program areas by taking advantage of existing tests and by targeting stubborn mutants that are hard to propagate. 

The studies of Papadakis and Malevris \cite{PapadakisM11, PapadakisM12} and Zhang et al. \cite{pexmutator} proposed embedding mutant related constraints, called infection conditions, within the meta-programs that inject and control the mutants in order to force symbolic execution to cover them. As a result, symbolic execution modules can produce test cases that satisfy the infection conditions and have good chances to kill the mutants. Although effective, these approaches only target mutant infection, which makes them relatively weak \cite{ChekamPTH17}.  

To bypasss the abovementioned problem, the studies of Papadakis and Malevris \cite{PapadakisM10} and Harman et al. \cite{HarmanJL11} aimed at indirectly handling mutant propagation. The former technique searches symbolically the path space of the mutant programs (after the mutation point), while the later one searches the input program space defined by the path conditions in order to bypass constraints not handled by the used solver and to indirectly make the mutants propagate. In contrast \tgtoolname aims at incrementally differentially exploring the path space by considering the symbolic states and making a relevant exploration. 

Fraser and Zeller \cite{FraserZ12} and Fraser and Arcuri \cite{FraserA15a} applied Search-based testing in order to generate mutation-based tests. Their key advancement was to guide the search by measuring the differences between the test traces of the original and mutant programs. While powerful, such an attempt still fails to provide the 
guidance needed in order to trigger such differences. 

Moreover, search techniques rely on the ability to execute test cases fast (applied at the unit level), making them less effective in cases of slow test execution (such as system level testing). Nevertheless, a comparison between search-based test generation and symbolic execution falls out of the scope of the present paper. 

Much of work on testing software patches has also been performed the recent years \cite{TanejaXTH11, MarinescuC12, MarinescuC13}. However, most of these methods aim at covering patches and not the program semantics (behavioural changes). Moreover, these techniques target the general patch testing problem, which in a sense assume very few patches with many changes. The mutation case though involves many mutants, which are small syntactic deviations, facts that our method takes advantage in order to optimize the mutant killings. 

Differential symbolic execution \cite{PersonDEP08} aims at reasoning about semantic differences of program versions, but since it performs a whole program analysis it can experience significant scalability issues when considering large programs and multiple mutants. Directed incremental symbolic execution \cite{PersonYRK11} guides the symbolic exploration through static program slicing. Unfortunately, such a method can be 
expensive when used with many mutants. Nevertheless, program slicing could be used to further guide \tgtoolname towards the relevant mutant exploration space. 


Shadow symbolic execution \cite{PalikarevaKC16} applies a combined execution on both program versions under analysis. It relies on analysis a meta-program that is similar to the mutant's meta-program in order to take advantage of the common program parts. The major difference with our method is that we specifically target multiple mutants at the same time, limit the program exploration through data state comparisons in order to optimize performance. Since shadow targets single patches and exhaustively searches the path space (after the mutation point) it can experience scalability issues. 

Overall, while many related techniques have been proposed, they have not been investigated in the context of mutation testing and particularly to target stubborn mutants. Stubborn mutants are hard to kill and their killing results in test inputs that are linked with corner cases and increase fault revelation \cite{ChekamPTH17}.


\section{Conclusion}  \label{sec:conclusion}


This paper introduced \tgtoolname, a method that generates test inputs for killing stubborn mutants. \tgtoolname relies on a form of shared differential symbolic execution that incrementally searches a small but `promising' code region around the mutation point in order to reveal divergent behaviours. This allows the fast and effective generation of test inputs that thoroughly exercise the targeted program corner cases. We have empirically evaluated \tgtoolname on Coreutils and demonstrated that it can kill approximately \changing{37}\% of the involved stubborn mutants within a two hour time budget. 
 
\bibliographystyle{ACM-Reference-Format}
\bibliography{sample-base}

\end{document}